\documentclass[runningheads,fleqn]{svmult}

\usepackage{makeidx}
\usepackage{graphicx}
\usepackage{multicol}
\usepackage{physmult}
\newcommand{\be}{\begin{eqnarray}}

\newcommand{\ee}{\end{eqnarray}}

\makeindex
\begin{document}
\title*{
         The Color Glass Condensate and Small x \\  Physics: 4 Lectures. }
\author{Larry McLerran
        \\
       {\small\it Nuclear Theory Group, Brookhaven National Laboratory,
        Upton, NY 11793  } \\      
       }
\authorrunning{Larry McLerran}
\maketitle
\parindent=20pt

\begin{abstract}

The Color Glass Condensate is a state of high density gluonic 
matter which controls the high energy limit of hadronic matter.  These 
lectures begin with a  discussion of 
general problems of high energy strong interactions.  The
infinite momentum frame description of a single hadron at very small x is
developed, and this picture is applied to the description of ultrarelativistic
nuclear collisions.  Recent developments in the renormalization group 
description of the Color Glass Condensate are reviewed. 
\end{abstract}

\section{Lecture I:  General Considerations}

\subsection{Introduction}

QCD is the correct theory of  hadronic physics.  It has been
tested in various experiments.  For high energy short distance
phenomena, perturbative QCD computations successfully confront
experiment.  In lattice Monte-Carlo computations, one gets a successful
semi-quantitative description of hadronic spectra, and perhaps in the
not too distant future one will obtain precise quantitative agreement.

\begin{figure}
\begin{center}
\includegraphics[width=0.5\textwidth] {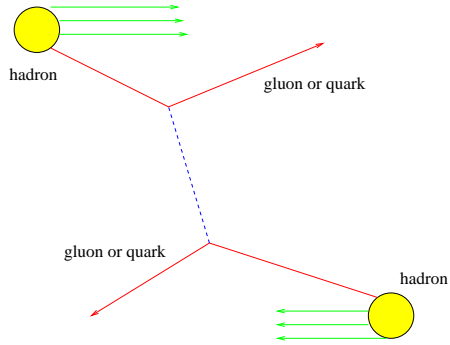}
\caption{Hadron-hadron scattering to produce a pair of jets. }
\label{jet}
\end{center}
\end{figure}

\begin{figure}
\begin{center}
\includegraphics[width=0.5\textwidth] {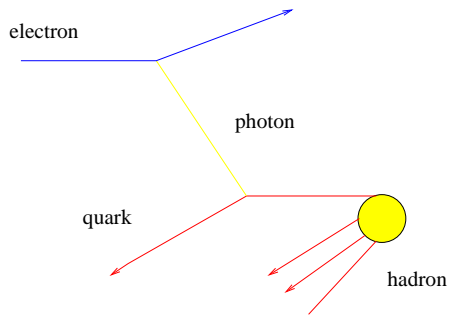}
\caption{Deep inelastic scattering of an electron on a hadron.}
\label{electron}
\end{center}
\end{figure}

At present, however, all analytic computations and all precise QCD tests
are limited to a small class of problems which correspond to short
distance physics, or to semi-quantitative comparisons with the results
of lattice gauge theory numerical computations.  
For the short distance phenomena, there is some characteristic energy transfer 
scale $E$, and one uses asymptotic freedom,
\be
        \alpha_S(E) \rightarrow 0
\ee
as $E \rightarrow \infty$
For example, in Fig. \ref{jet}, two hadrons collide to make a pair of
jets.  If the transverse momenta of the jets is large, the strong
coupling strength which controls this production is evaluated at the $p_T$ 
of the jet.  If $p_T >> \Lambda_{QCD}$, then the coupling is weak and this 
process can be computed in perturbation theory.
QCD has also been extensively tested in deep inelastic scattering.  In Fig. 
\ref{electron}, an electron exchanges a virtual photon with a hadronic target.
If the virtual photon momentum transfer $Q$ is large, then 
one can use weak coupling methods.

One question which we might ask is whether there are 
non-perturbative ``simple phenomena'' which arise from QCD which are
worthy of further effort.  The questions  I would ask before I would
become interested in understanding such phenomena are
\begin{itemize}

\item  Is the phenomenon simple in structure?

\item Is the phenomena pervasive?

\item  Is it reasonably plausible that one can understand the phenomena
from first principles, and  compute how it would appear in nature?

\end{itemize}

I will argue  that {\it gross} or
{\it typical} processes  in QCD,
which by their very nature are pervasive, 
appear to follow simple patterns.  The main content of this first lecture
is to show some of these processes, and pose some simple questions
about their nature which we do not yet understand.

My goal is to convince you that much of these average phenomena
of strong interactions at extremely high energies is controlled by a new form
of hadronic matter, a dense condensate of gluons.  This is called the 
Color Glass Condensate since
\begin{itemize}
\item Color:  The gluons are colored.
\item Glass:  We shall see that the fields associated with the glass 
evolve very slowly relative to natural time scales, and are disordered.  This 
is like a glass which is disordered and is a liquid on long time scales but
seems to be a solid on short time scales.
\item  Condensate:  There is a very high density of massless gluons.  These 
gluons can be packed until their phase space density is so high that 
interactions prevent more gluon occupation.  This forces at increasingly 
high  density
the gluons to occupy higher momenta, and the coupling becomes 
weak.  The density
saturates at $dN/d^2p_Td^2r_T \sim 1/\alpha_s >> 1$, and is a
condensate.
\end{itemize}
In these lectures, I will try to explain why the above is very plausible.

\subsection{Total Cross Sections at Asymptotic Energy}

Computing total cross sections as $E \rightarrow \infty$ is one of the
great unsolved problems of QCD.  
Unlike for processes which are computed in perturbation theory,
it is not required that any energy transfer become large as the total 
collision energy $E \rightarrow \infty$.  Computing a total cross section for 
hadronic scattering therefore appears to be intrinsically non-perturbative.
In the 60's and early 70's, Regge
theory was extensively developed in an attempt to understand the total
cross section.  The results of this analysis were to my mind
inconclusive, and certainly can not be claimed to be a first principles
understanding from QCD.

\begin{figure}
\begin{center}
\vspace{-1in}
\includegraphics[width=\textwidth] {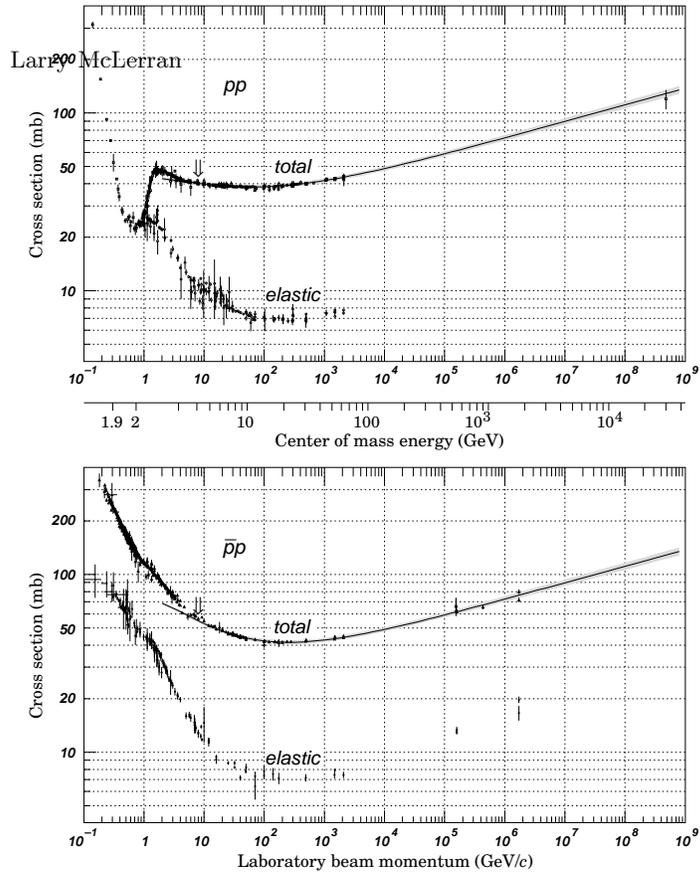}
\vspace{-1in}
\caption{The cross sections for $pp$ and $p\overline p$ scattering.}
\label{crosssection}
\end{center}
\end{figure}

The total cross section for  $pp$ and 
$\overline p p$ collisions is shown in Fig.
\ref{crosssection}.  
Typically, it is assumed that the total cross section grows as $ln^2
E$ as $E \rightarrow \infty$.  This is the so called Froisart bound
which corresponds to the maximal growth allowed by unitarity of the S
matrix.    Is this correct?  Is the coefficient of $ln^2 E$ universal
for all hadronic precesses?  Why is the unitarity limit saturated?  Can
we understand the total cross section from first principles in QCD?  Is
it understandable in weakly coupled QCD, or is it an intrinsically
non-perturbative phenomenon?

\subsection{How Are Particles Produced in High Energy Collisions?}

\begin{figure}
\begin{center}
\includegraphics[width=0.5\textwidth] {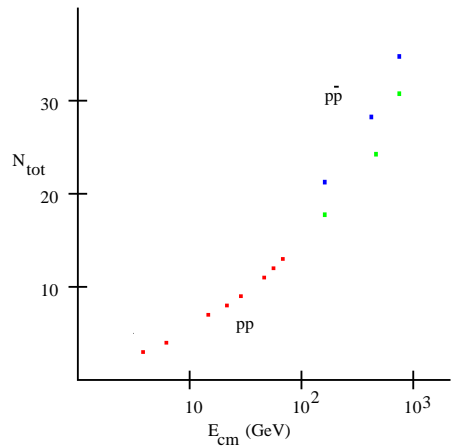}
\caption{The total multiplicity in $pp$ and $p \overline p$ collisions.}
\label{mult}
\end{center}
\end{figure}

In Fig. \ref{mult} , 
I plot the multiplicity of produced particles in $pp$ and in
$\overline p p$ collisions.  The last six  points correspond to
the $\overline p p$ collisions.  The three upper points
are the multiplicity in $p \overline p$ collisions, and the bottom
three have the mutliplicity at zero energy subtracted.
The remaining points correspond to $pp$.  Notice that the $pp$
points and those for $p \overline p$ with zero energy multiplicity subtracted
fall on the same curve. The implication is that whatever 
is causing the increase in multiplicity in these collisions may be from
the same mechanism.
Can we compute $N(E)$, the total multiplicity of
produced particles as a function of energy?

\subsection{Some Useful Variables}
At this point it is useful to develop some mathematical tools.  I will
introduce kinematic variables:  light cone coordinates.
Let the light cone longitudinal momenta be
\be
        p^\pm = {1 \over \sqrt{2}} (E \pm p_z)
\ee
Note that the invariant dot product
\be
        p \cdot q = p_t \cdot q_t - p^+q^- - p^- q^+
\ee
and that
\be
        p^+p^- = {1 \over 2} (E^2 - p_z^2) = {1 \over 2} (p_T^2 +m^2) = {1
\over 2} m_T^2 
\ee 
This equation defines the transverse mass $m_T$. 
(Please note that my metric is the negative of that conventionally used in
particle physics.) 

\begin{figure} 
\begin{center} 
\includegraphics[width=0.5\textwidth]
{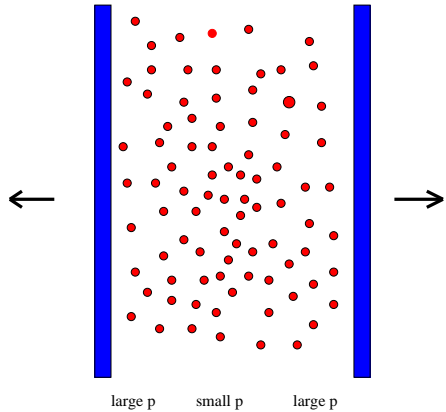} 
\caption{A hadron-hadron collision.  The produced particles are shown
as red circles. } 
\label{collision} 
\end{center}
\end{figure}

Consider a collision in the center of mass frame as shown in Fig. 
\ref{collision}.  In this figure, we have assumed that the colliding particles
are large compared to the size of the produced particles.  This is true for
nuclei, or if the typical transverse momenta of the produced particles is
large compared to $\Lambda_{QCD}$, since the corresponding size will be
much smaller than a Fermi.  We have also assumed that the colliding particles
have an energy which is large enough so that they pass through one another
and produce mesons in their wake.  This is known to happen experimentally:
the particles which carry the quantum numbers of the colliding particles
typically lose only some finite fraction of their momenta in the collision.

The right moving particle which initiates the collision shown in
Fig. \ref{collision} has $p_1^+ \sim \sqrt{2} \mid p_z \mid$ and
$p_1^- \sim {1 \over {2\sqrt{2}}} m_T^2/\mid p_z \mid$.
For the colliding particles $m_T = m_{projectile}$, that is because the
transverse momentum is zero, the transverse mass equals the particle
mass. For particle 2,
we have $p^+_2 = p^-_1$ and $p^-_2 = p^+_1$.

If we define the Feynman $x$ of a produced pion as
\be
        x = p^+_\pi /p_1^+
\ee
then $0 \le x \le 1$. (This definition agrees with Feynman's original one
if the energy of a particle in the center of mass frame is large and
the momentum is positive.  We will use this definition as a generalization
of the original one of Feynman since it is invariant under 
longitudinal Lorentz boosts.) 
The rapidity of a pion is defined to be
\be
        y = {1 \over 2} ln(p^+_\pi /p^-_\pi) = {1 \over 2} ln(2p^{+2}/m_T^2)
\ee
For pions, the transverse mass includes the transverse momentum of the
pion.

The pion rapidity is always in the range $-y_{CM} \le y \le y_{CM}$ where
$y_{CM} = ln({p^+/m_{projectile}})$  All the pions are produced in a
distribution of rapidities within this range.  

\begin{figure} 
\begin{center} 
\includegraphics[width=0.5\textwidth]
{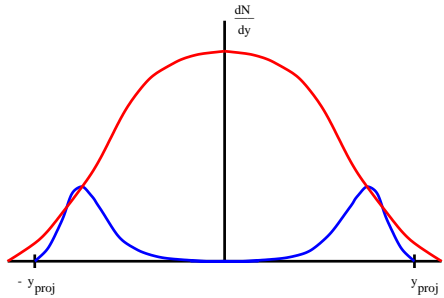} 
\caption{The rapidity distribution of particles produced in a hadronic 
collision. } 
\label{dndy} 
\end{center}
\end{figure}

A distribution of produced particles in a hadronic collision is shown in 
Fig. \ref{dndy}.  The leading particles are shown in blue and are clustered
around the projectile and target rapidities.  For example, in a heavy ion
collision, this is where the nucleons would be.  In red, the distribution
of produced mesons is shown.

These definitions are useful, among other reasons, because of their simple
properties under longitudinal Lorentz boosts:  $p^\pm \rightarrow
\kappa^{\pm 1} p^{\pm}$ where $\kappa$ is a constant.  Under boosts,
the rapidity just changes by a constant.  

The distribution of mesons, largely pions, shown in Fig. 4.  
are conveniently thought about  in
the center of mass frame.  Here we imagine the positive rapidity mesons as
somehow related to the right moving particle and the negative rapidity
particles as related to the left moving particles.  We define
$x = p^+/p^+_{projectile}$ and $x^\prime = p^-/p^-_{projectile}$
and use $x$ for positive rapidity pions and $x^\prime$ for negative
rapidity pions.

Several theoretical issues arise in multiparticle production.  Can we
compute $dN/dy$?  or even $dN/dy$ at $y = 0$?  How does the average transverse
momentum of produced particles $<p_T>$ behave with energy?  What is the
ratio of produced strange/nonstrange mesons, and corresponding ratios of
charm, top, bottom etc at $y = 0$ as the center of mass energy
approaches infinity?

Does multiparticle production as $E \rightarrow \infty$ at
$y = 0$ become simple, understandable and computable?

\begin{figure} 
\begin{center} 
\includegraphics[width=0.5\textwidth]
{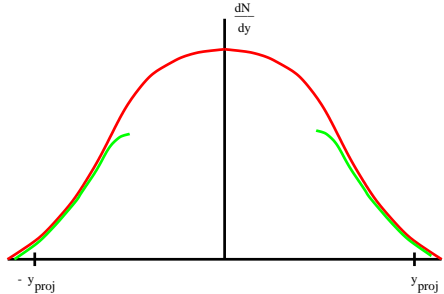} 
\caption{Feynman scaling of rapidity distributions. The two different colors
correspond to rapidity distributions at different energies.} 
\label{feynman} 
\end{center}
\end{figure}

There is a remarkable feature of rapidity distributions of produced 
\linebreak hadrons, 
which we shall refer to as Feynman scaling.  If we plot rapidity distributions
of produced hadrons at different energies, then as function of the distance 
from the fragmentation region, the rapidity distributions are to a good
approximation independent of energy.  This is illustrated in Fig. \ref{feynman}.
This means that as we go to higher and higher energies, the new physics is 
associated with the additional degrees of freedom at small rapidities
in the center of mass frame (small-x degrees of freedom).  
The large x degrees of freedom do not change much.
This suggests that there may be some sort of renormalization group description
in rapidity where the degrees of freedom at larger x are held fixed as we go to
smaller values of x.  We shall see that in fact these large x degrees of
freedom act as sources for the small x degrees of freedom, and the
renormalization group is generated by integrating out low x degrees of 
freedom to generate these sources.

\subsection{Deep Inelastic Scattering}

In Fig. \ref{electron},  deep inelastic scattering is shown.  Here an
electron emits a virtual photon which scatters 
from a quark in a hadron.  The momentum and energy
transfer of the electron is measured, and the results of the break up
are not.  In these lectures, I do not have sufficient time to
develop the theory of deep
inelastic scattering.  Suffice it to say, that this measurement is
sufficient at large momenta transfer $Q^2$ to measure the distributions
of quarks in a hadron.

To describe the quark distributions, it is convenient to work in a
reference frame where the hadron has a large longitudinal momentum
$p^+_{hadron}$.  The corresponding light cone momentum of the
constituent is $p^+_{constituent}$.  We define $x =
p^+_{constituent}/p^+_{hadron}$.  This x variable is equal to the
Bjorken x variable, which can be defined in a frame independent way.
In this frame independent definition, $x = Q^2/2p\cdot Q$ where $p$ is 
the momentum of the hadronic target and $Q$ is the momentum of the virtual
photon.  
The cross section which one extracts in deep inelastic scattering can be
related to the distributions of quarks inside a hadron, $dN/dx$.

It is useful to think about the distributions as a function of rapidity.
We define this for deep inelastic scattering as
\be
        y = y_{hadron} - ln(1/x)
\ee
and the invariant rapidity distribution as
\be
        dN/dy = x dN/dx
\ee
\begin{figure} 
\begin{center} 
\includegraphics[width=0.5\textwidth]
{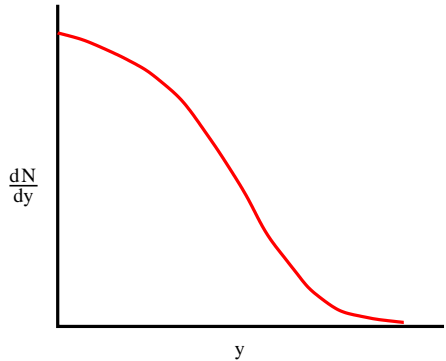} 
\caption{The rapidity distribution of gluons inside of a hadron. } 
\label{dndya} 
\end{center}
\end{figure}

In Fig. \ref{dndya},  
a typical $dN/dy$ distribution for  constituent gluons
of a hadron
is shown.  This plot is similar to the rapidity distribution of produced
particles in deep inelastic scattering.  The main difference is that we
have only half of the plot, corresponding to the left moving hadron in
a collision in the center of mass frame.

We shall later argue that there is in fact a relationship between the
structure functions as measured in deep inelastic scattering and the
rapidity distributions for particle production.  We will argue that the
gluon distribution function is in fact proportional to the pion rapidity
distribution.

The small x problem is that in experiments at Hera, the rapidity
distribution function for quarks grows as the rapidity difference between
the quark and the hadron grows.  This growth appears to be more rapid
than simply $\mid y_{proj} - y \mid$ or $(y_{proj}-y)^2$, 
and various theoretical models based on the
original considerations of Lipatov and colleagues suggest it may grow as
an exponential in $\mid y_{proj} - y \mid$.\cite{bfkl} (Consistency of
the BFKL approach with the more established DGLAP evolution equations
remains an outstanding theoretical problem.\cite{dglap})  
If the rapidity distribution grew at most as
$y^2$, then there would be no small x problem.
We shall try to explain
the reasons for this later in this lecture.
\begin{figure} 
\begin{center} 
\includegraphics[width=0.5\textwidth]
{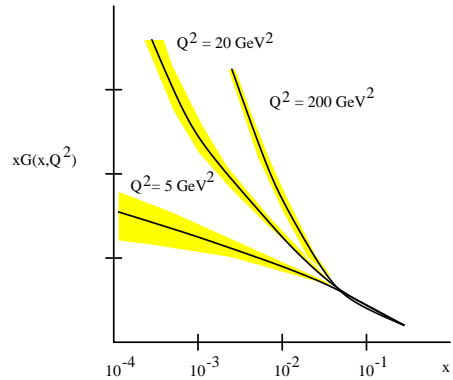} 
\caption{The Zeus data for the gluon structure functions. } 
\label{gluon} 
\end{center}
\end{figure}

In Fig. \ref{gluon}, the Zeus data for the gluon structure function is
shown.\cite{z} I have plotted the structure function for $Q^2 = 5~ GeV^2$,
$20~ GeV^2$ and $200~GeV^2$.  
The structure function depends upon the resolution of
the probe, that is $Q^2$.  Note the rise of $xg(x)$ at small x, this is
the small x problem.  If one had plotted the total multiplicity of
produced particles in $pp$ and $\overline p p$ collisions on the same plot,
one would have found rough agreement in the shape of the curves.  
Here I would  use $y = log(E_{cm}/1~GeV)$
for the pion production data.  This is approximately the maximal value of
rapidity difference between centrally produced pions and the projectile
rapidity.  The total multiplicity would be rescaled so that at small x, it
matches the gluon structure functions.    This demonstrates the qualitative
similarity between the gluon structure function and the total
multiplicity. 
\begin{figure} 
\begin{center} 
\includegraphics[width=0.5\textwidth]
{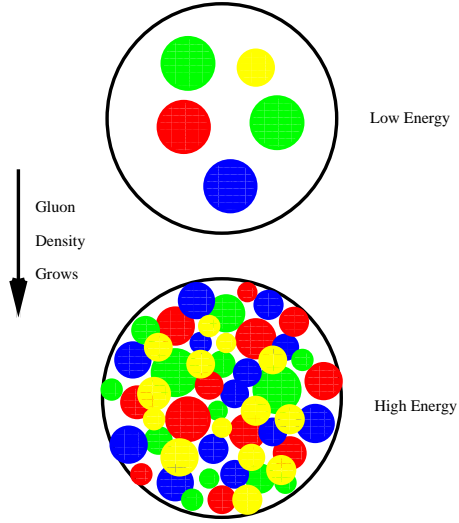} 
\caption{Saturation of gluons in a hadron.  A view of a hadron head on as
x decreases.} 
\label{saturation} 
\end{center}
\end{figure}

Why is the small x rise in the gluon distribution a problem?  
Consider
Fig. \ref{saturation}, 
where we view hadron head on.\cite{glr}-\cite{mq} The constituents are the
valence quarks, gluons and sea quarks
shown as colored circles.  As we add more and more constituents,
the hadron becomes more and more crowded.  If we were to try to measure
these constituents with say an elementary photon probe, as we do in deep
inelastic scattering, we might expect that the hadron would become so
crowded that we could not ignore the shadowing effects of constituents as
we make the measurement.  (Shadowing means that some of the partons are
obscured by virtue of having another parton in front of them.  For hard
spheres, for example, this would result in a decrease of the scattering
cross section relative to what is expected from incoherent independent
scattering.) 

In fact, in deep inelastic scattering, we are measuring the cross section
for a virtual photon $\gamma^*$ and a hadron, $\sigma_{\gamma^*
hadron}$.
Making x smaller corresponds to increasing the energy of the interaction
(at fixed $Q^2$).  An exponential growth in the rapidity corresponds to
power law growth in $1/x$, which in turn implies power law growth with
energy.  This growth, if it continues forever, violates unitarity.  The
Froissart bound will allow at most $ ln^2 (1/x)$.  (The Froissart
bound is a limit on how rapidly a total cross section can rise.  It
follows from the unitarity of the scattering matrix.)

We shall later argue that in fact the distribution functions at fixed
$Q^2$ do in fact saturate and cease growing so rapidly at high energy.
The total number of gluons however demands a resolution scale, and we
will see that the natural intrinsic scale is growing at smaller values
of x, so that effectively, the total number of gluons within this
intrinsic scale is always increasing.  The quantity
\be
        \Lambda^2 = {1 \over {\pi R^2}} {{dN} \over {dy}} 
\ee 
defines this
intrinsic scale.  Here $\pi R^2$ is the cross section for hadronic
scattering from the hadron.  For a nucleus, this is well defined.  For a
hadron, this is less certain, but certainly if the wavelengths of probes
are small compared to $R$, this should be well defined.  If \be
        \Lambda^2  >> \Lambda^2_{QCD}
\ee
as the Hera data suggests, then we are dealing with weakly coupled QCD
since $\alpha_S(\Lambda) << 1$.

Even though QCD may be weakly coupled at small x, that does not mean the
physics is perturbative.  There are many examples of nonperturbative
physics at weak coupling.  An example is instantons in electroweak
theory, which lead to the violation of baryon number.   Another example
is the atomic physics of highly charged nuclei.  The electron propagates
in the background of a strong nuclear Coulomb field, but on the other
hand, the theory is weakly coupled and there is a systematic weak
coupling expansion which allows for computation of the
properties of high Z (Z is the charge of the nucleus) atoms.

We call this assortment of gluons a Color Glass Condensate.
The name follows from the fact that the gluons are colored,
and we have seen that they are very dense. For massless particles
we expect that the high density limit will be a Bose condensate.
The phase space density will be limited by repulsive gluon interactions,
and be of order $1/\alpha_s >> 1$.  The glass nature follows because
these fields are produced by partons at higher rapidity, and in the 
center of mass frame, they are Lorentz time dilated.  Therefore the induced
fields at smaller rapidity evolve slowly compared to natural time scales.
These fields are also disordered.  These two properties are similar 
to that of a glass which is a disrodered material which is a
liquid on long time scales and a solid on short ones.

If the theory is local in rapidity, then the only parameter which can
determine the physics at that rapidity is $\Lambda^2$.
Locality in rapidity means that there are not long range correlations
in the hadronic wavefunction as a function of rapidity.  In pion production,
it is known that except for overall global conserved quantities such as 
energy and total charge, such 
correlations are of short range.
Note that if only $\Lambda^2$ determines the physics, then in an
approximately scale invariant theory such as QCD,  a typical transverse
momentum of a constituent will also be of order $\Lambda^2$.  If
$\Lambda^2 >> 1/R^2$, where $R$ is the radius of the hadron, then
the finite size of the hadron becomes irrelevant.  Therefore at small
enough x, all hadrons become the same.  The physics should only be
controlled by $\Lambda^2$.

There should therefore be some equivalence between nuclei and say
protons.  When their $\Lambda^2$ values are the same, their physics
should be the same.  We can take an empirical parameterization of the
gluon structure functions as
\be
        {1 \over {\pi R^2}} {{dN} \over {dy}} \sim {{A^{1/3}} \over x^\delta}
\ee
where $\delta \sim .2 - .3$.  This suggests that there should be the
following correspondences:
\begin{itemize}

\item RHIC with nuclei $\sim$ Hera with protons

\item LHC with nuclei $\sim$ Hera with nuclei

\end{itemize}

Estimates of the parameter $\Lambda$ for nuclei at RHIC energies give
$\sim 1-2 ~Gev$, and at LHC $2-3 ~Gev$.

Since the physics of high gluon density is weak coupling we have the
hope that we might be able to do a first principle calculation of
\begin{itemize}

\item the gluon distribution function

\item the quark and heavy quark distribution functions

\item the intrinsic $p_T$ distributions quarks and gluons

\end{itemize}

We can also suggest a simple escape from unitarity arguments which
suggest that the gluon distribution function must not grow at
arbitrarily small x.  The point is that at smaller x, we have larger
$\Lambda$ and correspondingly larger $p_T$.  A typical parton added to
the hadron has a size of order $1/p_T$.  Therefore although we are
increasing the number of gluons, we do it by adding in more gluons of
smaller and smaller size.  A probe of size resolution $\Delta x \ge
1/p_T$  at fixed $Q$ will not see partons smaller than this resolution
size.  They therefore do not contribute to the fixed $Q^2$ cross
section, and there is no contradiction with unitarity.

\subsection{Heavy Ion Collisions}

\begin{figure} 
\begin{center} 
\includegraphics[width=0.5\textwidth]
{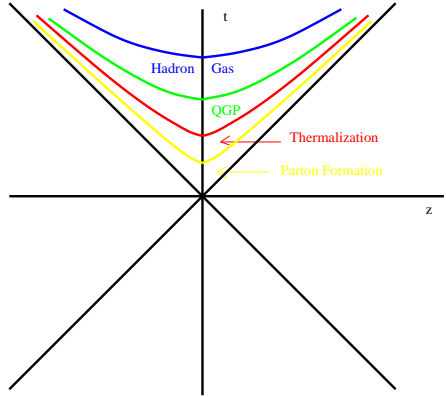} 
\caption{A space-time figure for ultrarelativistic heavy ion collisions. } 
\label{spacetime} 
\end{center}
\end{figure}

In Fig. \ref{spacetime}, the standard 
lightcone cartoon of heavy ion collisions is
shown.\cite{bj}  
To understand the figure, imagine we have two Lorentz contracted
nuclei approaching one another at the speed of light.  Since they are
well localized, they can be thought of as sitting at $x^\pm = 0$, that
is along the light cone, for $t < 0$.  At $x^\pm = 0$, the nuclei
collide.  To analyze this problem for $t \ge 0$, it is convenient to
introduce a time variable which is Lorentz covariant under longitudinal
boosts
\be
        \tau = \sqrt{t^2 - z^2}
\ee
and a space-time rapidity variable
\be
        \eta = {1 \over 2} ln\left( {{t-z} \over {t+z}} \right)
\ee
For free streaming particles
\be
        z = vt = {p_z \over E} t
\ee
we see that the space-time rapidity equals the momentum space rapidity
\be
        \eta = y
\ee

If we have distributions of particles which are slowly varying in
rapidity, it should be a good approximation to take the distributions to
be rapidity invariant.  This should be valid at very high energies in
the central region.  By the correspondence between space-time and
momentum space rapidity, it is plausible therefore to assume  that
distributions are independent of $\eta$.  Therefore distributions are
the same on lines of constant $\tau$, which is as shown in 
Fig. \ref{spacetime}.
At $z = 0$, $\tau = t$, so that $\tau $ is a longitudinally Lorentz
invariant time variable.

We expect that at very late times, we have a free streaming gas of
hadrons.  These are the hadrons which eventually arrive at our detector.
At some earlier time, these particles decouple from a dense gas of
strongly interacting hadrons.  As we proceed earlier in time, at some
time there is a transition between a gas of hadrons and a plasma of
quarks and gluons.  This may be through a first order phase transition
where the system might exist in a mixed phase for some length of time,
or perhaps there is a continuous change in the properties of the system

At some earlier time, the quarks and gluons of the quark-gluon plasma
are formed.  This is at RHIC energies, a 
time of the order of a Fermi, perhaps as
small as $.1 ~Fermi$.  As they form, the particles scatter from one
another, and this can be described using the methods of transport
theory.  At some later time they have thermalized, and the system can be
approximately described using the methods of perfect fluid
hydrodynamics.

In the time between that for which the quarks and gluons have been
formed and $\tau = 0$, the particles are being formed.  This is where
the initial conditions for a hydrodynamic description are made.

In various levels of sophistication, one can compute the properties of
matter made in heavy ion collisions at times later than the formation
time.  The problems are understood in principle for $\tau \ge
\tau_{formation}$ if perhaps not in fact.
Very little is known about the initial conditions.

In principal, understanding the initial conditions should be the
simplest part of the problem.  At the initial time, the degrees of
freedom are most energetic and therefore one has the best chance to
understand them using weak coupling methods in QCD.

There are two separate classes of problems one has to understand for the
initial conditions.  First the two nuclei which are colliding are in
single quantum mechanical states.  Therefore for some early time, the
degrees of freedom must be quantum mechanical.  This means that
\be
        \Delta z \Delta p_z \ge 1
\ee
Therefore classical transport theory cannot describe the particle down
to $\tau = 0$ since classical transport theory assumes we know a
distribution function $f(\vec{p}, \vec{x}, t)$, which is a simultaneous
function of momenta and coordinates.  This can also be understood
as a consequence of entropy.  An initial quantum state has zero entropy.
Once one describes things by classical distribution functions,
entropy has been produced.  Where did it come from?

Another problem which must be understood is classical charge coherence.
At very early time, we have a tremendously large number of particles
packed into a longitudinal size scale of less than a fermi.  This is due
to the Lorentz contraction of the nuclei.  We know that the particles
cannot interact incoherently.  For example, if we measure the field due
to two opposite charge at a distance scale $r$ large compared to their
separation, we know the field falls as $1/r^2$, not $1/r$.  On the other
hand, in cascade theory, interactions are taken into account by cross
sections which involve matrix elements squared.  There is no room for
classical charge coherence.

There are a whole variety of problems one can address in heavy ion
collisions such
\begin{itemize}

\item What is the equation of state of strongly interacting matter?

\item Is there a first order QCD phase transition?

\end{itemize}
These issues and others would take us beyond the scope of these
lectures.  The issues which I would like to address are related to the
determination of the initial conditions, a problem which can hopefully
be addressed using weak coupling methods in QCD.

\subsection{Universality}

There are two separate formulations of universality which are important
in understanding small x physics.

The first is a weak universality.  This is the statement that physics
should only depend upon the variable\cite{mv}
\be
        \Lambda^2 = {1 \over {\pi R^2}} {{dN} \over {dy}}
\ee
As discussed above, this universality has immediate experimental\linebreak
consequences which can be directly tested.

The second is a strong universality which is meant in a statistical
mechanical sense.  At first sight it appears to be  a formal idea with little
relation to experiment.  If it is however true, its consequences are
very powerful and far reaching.  What we shall mean by strong
universality is that the effective action which describes small x
distribution function is critical and at a fixed point of some
renormalization group.  This means that the behavior of correlation
functions is given by universal critical exponents, and these universal
critical exponents depend only on general properties of the theory such
as the symmetries and dimensionality.

Since the correlation functions determine the physics, this statement
says that the physics is not determined by the details of the
interactions, only by very general properties of the underlying theory!

\section{Lecture II: A Very High Energy Nucleus}

In this lecture, I will consider the properties of a 
single nucleus.\cite{mv}-\cite{k}
I will develop the theory of the small x part of the 
nucleus, the components
most relevant in the high energy limit.  I will begin with some general
considerations.  This will 
largely be done to develop approximations which will be useful later,
and leads directly to the Color Glass description.  I then
present a brief review of light cone quantization.  Finally, I turn to a 
computation of the color fields which describe the nuclear wavefunction at 
small x.  I show that in a simple approximation for the Color Glass,
one recovers saturation, and that the phase space density of the fields is of 
order $1/\alpha_s$ which is typical of a Bose condensate.  

\subsection{Approximations and the Color Glass}

In the previous lecture, I argued that when we go to small x 
\be
	\Lambda^2 = {1 \over {\pi R^2}} {{dN} \over {dy}} >> \Lambda_{QCD}^2
\ee
that the theory is weakly coupled $\alpha_s(\Lambda) << 1$.  The typical 
transverse momentum scale of constituents of this low x part of the hadron 
wavefunction is
\be 
	p_T^2 \sim \Lambda^2 >> 1/R_{had}^2
\ee
This equation means that the scale of transverse variation of the hadron
is over much larger sizes than the transverse De Broglie wavelength.
I can therefore treat the hadron as having a well defined size and collisions
will have well defined impact parameter.   

For our purposes, it is
sufficient to treat the hadron as a thin sheet of infinite transverse extent.
The transverse variation in radius can be reinserted in an almost
trivial generalization of these considerations.  The thinness of the sheet
follows because I shall assume that the sources for the fields at small x
come from partons at much larger $x$ which are Lorentz contracted to size 
scales much smaller than can be resolved.  
\begin{figure} 
\begin{center} 
\includegraphics[width=0.5\textwidth]
{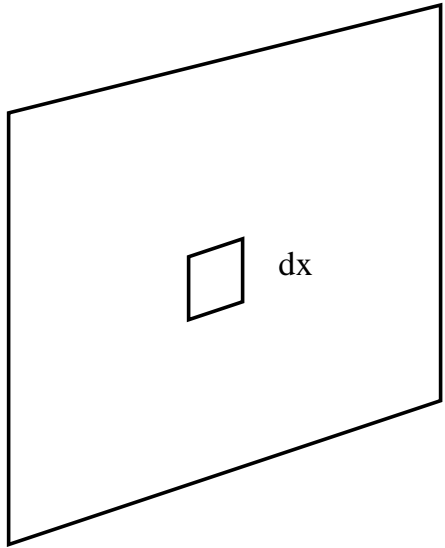} 
\caption{A single nucleus in the infinite momentum frame as seen by a small
x probe. } 
\label{sheet} 
\end{center}
\end{figure}
In Fig. \ref{sheet}, a nucleus in the infinite momentum frame is shown,
within the approximations described above.

Recall from the first lecture, we introduced rapidities associated
with produced particles in hadron-hadron collisions,
\be
	y & =  & {1 \over 2} ln(p^+/p^-)  =  
  ln(\sqrt{2}p^+/M_T)  \nonumber \\ & = & 
           ln(\sqrt{2}p^+_{had}/M_T) + ln(p^+/p^+_{had})  
\sim y_{had} - ln(1/x)
\ee
This expression shows that the rapidity of produced hadrons can be written in 
the form used to describe the rapidity of constituents of a hadron.  If
we were to think of both the constituents and produced partons as 
pions, they would be the same, 
or alternatively if we think of both the produced and constituent 
partons as gluons.  We can 
convert to spacetime rapidity using the uncertainty relation $p^\pm x^\mp 
\sim 1$.
\be
	y \sim {1\over 2}ln(x^-/x^+) \sim y_{had} - ln(x^-p^+_{had})
\ee
We have assumed in deriving this relationship that the typical values
of the proper time $\tau = \sqrt{x^+x^-}$ are not large compared to natural 
scales such as a transverse mass.
These relations argue that all rapidities, up to shifts of order one, 
are the same.  We can identify all momentum-space and space-time rapidities!
This has the profound consequence that at high energies momentum space and 
space-time are intrinsically correlated, and particles which arise from a 
localized region of momentum space rapidity also arise from a localized 
region of space-time rapidity.

\begin{figure} 
\begin{center} 
\includegraphics[width=0.5\textwidth]
{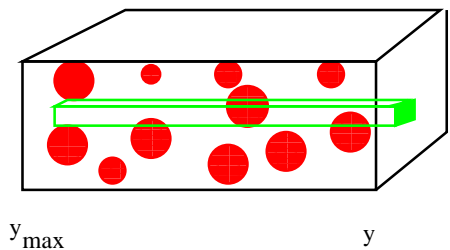} 
\caption{A single nucleus shown in terms of the space-time rapidity.  
The red circles indicate partons. } 
\label{intersect} 
\end{center}
\end{figure}

Now we illustrate a high energy hadron in terms of space-time rapidity.
This has the effect of spreading out the thin sheet shown in Fig. \ref{sheet},
as shown in Fig. \ref{intersect}.  Note that the partons which are
shown  have an ordering in momentum space rapidity which corresponds
to their coordinate space rapidity.  Fast partons are to the left.
In this Figure, I have drawn a tube
of transverse extent $dx$ which goes through the nucleus.  I take 
$dx << 1~Fm$ so that one is resolving the constituents of ordinary hadrons.
Notice that when $dx \rightarrow 0$, the longitudinal separation between
hadrons which intersect the tube becomes large.  If I also require that the 
energy is high enough so that there is always a large number of partons
which intersect the box (which are longitudinally well separated), then 
one can think of a source associated with the charge inside the box,
and this charge has a random distribution over the transverse area of the box.
(At what scale the source becomes random is not entirely clear 
from this discussion. This will be resolved more carefully later.)

In the limit where $ 1/\Lambda << dx << 1 ~Fm$, there are many charges
inside the box of dimension $dx$.  The charge should go over to a classical 
charge on this resolution scale because we can ignore commutators of charges
\be
	 \mid [Q^a, Q^b] \mid  = \mid i f^{abc} Q^c \mid << Q^2
\ee
We can define a current associated with this charge which is localized in 
the sheet as
\be
	J^\mu_a = \delta^{\mu +} \delta (x^-) q^a(x_T)
\ee
The $+$ component of the current is the only important one because
the sheet is traveling near the speed of light.  The source $q^a(x_T)$
is a c-number color charge density which is a random variable on the sheet.
It is only defined on scales  $1/\Lambda << dx << 1 ~ Fm$.  The delta 
function of $x^-$ expresses the fact that the source is on  a thin sheet.
In fact, for many applications, we will have to relax the delta function
assumption, and work with  a charge density which includes the effect of
distribution in $x^-$ as
\be
	q^a(x_T) = \int dx^- ~\rho^a (x^-,x_T)
\ee
and where for many purposes
\be
	\rho^a(x^-, x_T) \sim \delta (x^-) q^a(x_T)
\ee

We now know how to write down a theory.  It is a theory where one computes
the classical gluon field in stationary phase approximation and then integrates
over a random source function.  Its measure is
\be
	Z = \int [dA] [d\rho] exp\left\{iS[A] + iJ^+A^- - {1 \over 2}
\int dx^- d^2x_T {{\rho^2(x^-,x_T)} \over {\mu^2(x^-)}}\right\}
\ee
In this theory, we have assumed that the sources are randomly distributed as 
a Gaussian.  This turns out to be an approximation valid in a particular
range of resolution $dx$, and can be fixed up for a wider range.  This will
be discussed when we do the renormalization group.  The sources and fields 
are coupled together in the standard $J \cdot A$ form.  This results in the
problem that the extended current conservation law $D_\mu J^\mu = 0$ makes 
$J$ not an independent function.  This problem can be avoided by
introducing a
generalization of the $J \cdot A$ coupling.  This generalization
turns out to be important for
the renormalization group analysis of this theory, but is not important when
we compute the classical field associated with these sources.

The above theory implicitly has cutoffs in it.  We have discussed the range of
$dx$ for which this effective theory is valid.  Implicit in the analysis is 
that the fields we are computing have $p^+$ values much less 
than those of the sources.
This implies there is an upper $p^+$ cutoff in the fields $A$ considered.
If we were to Fourier analyze the sources $\rho$, they would have their support
for $\mid p^+ \mid$ which is greater than that of the cutoff.  This cutoff
is of course entirely arbitrary, and the lack of dependence of physical
quantities upon this cutoff forms the basis of the renormalization group.

Notice that this theory, in spite of having a gauge dependent 
source, is gauge invariant on account of the integration over all sources.
This computation of classical fields associated with sources and then averaging
over sources is similar to the mathematics of glasses.  The physical 
origin of this similarity is the Lorentz time
dilation of the source for the fields and the disorder
of the gluon field.  The Lorentz time dilation is of course an 
approximation, and if one were to observe these classical fields over 
long enough time scales they would evolve, as do the atoms in a glass.

Notice that
\be
	< \rho^a(x) \rho^b(y) > = \delta^{ab} \delta^{(3)} (x-y) \mu^2 (x^-)
\ee
so that $\mu^2$ is the charge squared per unit transverse area per unit $x^-$
scaled by $1/(N_c^2-1)$.

\subsection{Light Cone Quantization}

Before discussing the properties of classical fields associated 
with these \linebreak
sources, it is useful to review some properties of light cone 
quantization.\cite{v}
This will allow us to pick out physical observables, such as the gluon density,
from expectation values of gluon field operators.

Light cone coordinates are
\be
        x^\pm = {1 \over \sqrt{2}} (x^0 \pm x^3)
\ee
and momenta
\be
        p^\pm = {1 \over \sqrt{2}} (p^0 \pm p^3)
\ee
The invariant dot product is
\be
        p \cdot x = p_t \cdot x_t -p^+x^--p^-x^+
\ee
where $p_t$ and $x_t$ are transverse coordinates.  This implies that in
this basis the metric is $g^{+-} = g^{-+} = -1$, $g^{ij} = \delta^{ij}$
where $i,j$ refer to transverse coordinates.  All other elements of the
metric vanish.

An advantage of light cone coordinates is that if we do a Lorentz boost
along the longitudinal direction with Lorentz gamma factor $\gamma =
cosh(y)$ then $p^\pm \rightarrow e^{\pm y} p^\pm$

\begin{figure} 
\begin{center} 
\includegraphics[width=0.5\textwidth]
{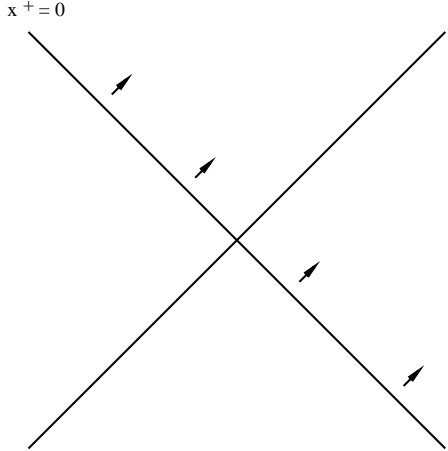} 
\caption{The initial value problem in light cone coordinates. } 
\label{lightconein} 
\end{center}
\end{figure}

If we let $x^+$ be a time variable, we see that the variable $p^-$ is to
be interpreted as an energy.  Therefore when we have a field theory, the
component of the momentum operator $P^-$ will be interpreted as the
Hamiltonian.  The remaining variables are to be thought of as momenta and
spatial coordinates.  In Fig. \ref{lightconein}, 
there is a plot of the $z,t$ plane.
The line $x^+ =0$ provides a surface where initial data might be
specified.  Time evolution is in the direction normal to this surface.

We see that an elementary wave equation
\be
        (p^2 + M^2) \phi = 0
\ee
is particularly simple in light cone gauge.  Since $p^2 = p_t^2 - 2
p^+p^-$ this equation is of the form
\be
        p^- \phi = {{p_t^2 + M^2} \over {2p^+}} \phi
\ee
is first order in time.  In light cone coordinates, the dynamics looks
similar to that of the Schrodinger equation.  The initial data to be
specified is only the value of the field on the initial surface.

In the conventional treatment of the Klein-Gordon field, one must
specify
the field and its first derivative (the momentum) on the initial
surface.  In  light cone coordinates, the field is sufficient and the
field momentum is redundant.  This means that the field momentum will
not commute with the field on the initial time surface!

Lets us work all this out with the example of the Klein Gordon field.
The action for this theory is
\be
        S = - \int d^4x \left\{ {1 \over 2} (\partial \phi )^2 + {1 \over 2}
M^2 \phi^2 \right\}
\ee
The field momentum is
\be
        \Pi (x_t, x^-) = {{\delta S} \over {\delta \partial_+ \phi}} =
\partial_- \phi = {\partial \over {\partial x^-}} \phi
\ee
Note that $\Pi$ is a derivative of $\phi$ on the initial time surface.
It is therefore not an independent variable, as would be the case in the
standard canonical quantization of the scalar field.

We postulate the equal time commutation relation
\be
        [ \Pi(x_t,x^-) , \phi(y_t,y^-)] = -{i \over 2} \delta^{(3)} (x-y)
\ee
(The factor of two in the above expression is subtle and comes from a careful
reduction of constrained Dirac bracket quantization for the classical theory
to quantum field theory.  It can be checked by verifying that we get
the correct result for the Hamiltonian.)
Here the time is $x^+ = y^+ = 0$ in both the the field and field
momentum.  We see therefore that
\be
        \partial_- [\phi(\vec{x}), \phi(\vec{y}) ] = -{i \over 2}
 \delta^{(3)} (x-y)
\ee
or
\be
        [\phi (x), \phi(y)] = -{ i \over 2} 
\epsilon(x^- - y^-) \delta^{(2)} (x-y)
\ee
Here $\epsilon (v)$ is $1/2$ for $v > 0$ and $-1/2$ for $v < 0$.

These commutation relations may be realized by the field
\be
        \phi(x) & = & \int {{d^3p} \over {(2\pi)^3 2\sqrt{2}p^+}} 
e^{ipx} a(p)  
\nonumber  \\
                         & = & \int_{p^+ > 0} {{d^3p} \over
{(2\pi)^32\sqrt{2}p^+}} 
\left\{ e^{ipx} a(p) + e^{-ipx} a^\dagger (p) \right\}
\ee
Using
\be
        [ a(p), a^\dagger (q) ] = 2 p^+ (2\pi )^3 \delta^{(3)} (p-q)
\ee
one can verify that the equal time commutation relations for the
field are satisfied.

The quantity $1/p^+$ in the expression for the field in terms of
creation and annihilation operators is singular when $p^+ = 0$.  When we
use a principle value prescription, we reproduce the form of the
commutation relations postulated above with the factor of $\epsilon
(x^- - y^-)$.  Different prescriptions correspond to different choices
for the inversion of $1 \over \partial ^-$.  One possible prescription
is the Leibbrandt-Mandelstam prescription $1/p^+ = p^-/(p^+p^- +
i\epsilon)$.  This prescription has some advantages relative to the
principle value prescription in that it maintains causality at
intermediate stages of computations and the principle value prescription
does not.  In the end, for physical quantities, the choice of
prescription cannot result in different results.  Of course, in some
schemes the computations may become prohibitively difficult.

The light cone
Hamiltonian is
\be
        P^- = \int_{p^+ > 0 } {{d^3p} \over {(2\pi)^3 2p^+}}
{{p_t^2 + M^2} \over {2p^+}} a^\dagger (p) a(p)
\ee
with obvious physical interpretation.

In a general interacting theory, the Hamiltonian will of course be more
complicated.  The representation for the fields in terms of creation and
annihilation operators will be the same as above.  Note that all
particles created by a creation operator have positive $P^+$.
Therefore, since the vacuum has $P^+ = 0$, there can be no particle
content to the vacuum.  It is a trivial state.  Of course this must be
wrong since the physical vacuum must contain condensates such as the one
responsible for chiral symmetry restoration.  It can be shown that such
non-perturbative condensates arise in the $P^+ = 0$ modes of the theory.
We have not been careful in treating such modes.  For perturbation
theory, presumably to all orders, the above treatment is sufficient for
our purposes.

\subsection{Light Cone Gauge QCD}

In QCD we have a vector field $A^\mu_a$.  This can be decomposed into
longitudinal and transverse parts as
\be
        A^\pm_a = {1 \over \sqrt{2}} (A^0_a \pm A^z_a)
\ee
and the transverse as lying in the two dimensional plane orthogonal to
the beam z axis.  Light cone gauge is
\be
        A^+_a = 0
\ee

In this gauge, the equation of motion
\be
        D_\mu F^{\mu \nu} = 0
\ee
is for the $+$ component
\be
        D_iF^{i+} - D^+F^{-+} = 0
\ee
which allows one to compute $A^-$ in terms of $A^i$ as
\be
        A^- = {1 \over \partial^{+2}} D^i \partial^+ A^i
\ee
This equation says that we can express the longitudinal field entirely
in terms of the transverse degrees of freedom which are specified by
the transverse fields entirely and explicitly.  These degrees of freedom
correspond to the two polarization states of the gluons.

We therefore have
\be
        A^i_a (x) = \int_{p^+ > 0} {{d^3p} \over {(2\pi)^3 2p^+}} \left(
e^{ipx} a^i_a(p) + e^{-ipx} a^{i\dagger}_a (p)\right)
\ee
where
\be
        [a^i_a (p), a^{j\dagger}_b(q)] = 2p^+ \delta_{ab} \delta^{ij} (2\pi)^3
\delta^{(3)} (p - q)
\ee
where the commutator is at equal light cone time $x^+$.

\subsection{Distribution Functions}

We would like to explore some hadronic properties using light cone field
operators.  For example, suppose we have a hadron and ask what is the
gluon content of that hadron.  Then we would compute
\be
        {{dN_{gluon}} \over {d^3p}} = <h\mid a^\dagger (p) a(p) \mid h>
\ee
If we express this in terms of the gluon field, we find
\be
        {{dN_{gluon}} \over {d^3p}} = {{2p^+} \over {(2\pi)^3}}
<h \mid A^{ia}(\vec{p},x^+) A^{ia} (-\vec{p},x^+) \mid h >
\ee
which can be related to the gluon propagator. 	
The quark distribution for quarks of flavor $i$ (for the sum of quarks
and antiquarks) would be given in terms of creation and annihilation
operators for quarks as
\be
        {{dN_i} \over {d^3p}} = <h \mid \{b_i^\dagger (p) b_i(p) + d_i^\dagger
(p) d_i(p) \} \mid  h>
\ee
where $b$ corresponds to quarks and $d$ to antiquarks.
The creation and annihilation operators for quarks and gluons can be
related to the quark coordinate space field operators by techniques
similar to those above.\cite{v1}

\subsection{The Classical Gluon Field}

To compute the gluon distribution function, we need the expectation
value of the gluon field.  To lowest order in weak coupling, this is given
by computing the classical gluon field and then averaging over sources.
The classical equation of motion is
\be
	D_\mu F^{\mu \nu } = \delta^{\nu +} \rho (x^-, x_T)
\ee

To solve this equation we shall work in the gauge ${\cal A}^- = 0$, and then 
gauge rotate the solution back to lightcone gauge $A^+ = 0$.
The solution in ${\cal A}^- = 0 $ gauge is 
\be
	{\cal A}^i & = & 0 \nonumber \\
         - \nabla_T^2 {\cal A}^+ & =  & {\overline \rho}
\ee
Here ${\overline\rho} = U^\dagger (x) \rho U(x)$ is the source which has been
gauge rotated to this new gauge.  Since the measure for integration over
sources is gauge invariant, we do not have to distinguish between these
sources since we can rotate one into the other.

To rotate back to lightcone gauge we use
\be
	{\cal A}^\mu = U^\dagger A^\mu U + 
{i \over g} U^\dagger \partial^\mu U
\ee
so that the gauge rotation matrix $U$ is
\be
	\partial^+ U = -ig U {\cal A}^+
\ee
where
\be
	{\cal A}^+ = \alpha = {1 \over {-\nabla_T^2}} \nabla {\overline \rho}
\ee
The solution is\cite{k}-\cite{mkmw}
\be
	U^\dagger = P exp \left\{ ig \int^{x^-}_{x_0^-} dz^- 
\alpha (z^-,x_T)\right\}
\ee
There is a choice of boundary condition here associated with $x_0^+$.
The ambiguity with this choice is associated with a residual gauge freedom.
We shall resolve this by choosing retarded boundary conditions, 
$x_0^- \rightarrow - \infty$.  This boundary condition lets us construct
the solution for $U$ at some $x^-_1$ knowing only information about
$\alpha$ for  $x^- < x^-_1$.  

The solution in light cone gauge is therefore
\be
	A^+ & =  & A^- = 0 \nonumber \\
        A^i  & = & {i \over g} U \nabla^i U^\dagger
\ee
If $x^-$ is outside of the range of support of the source $\rho$, this can 
be written as 
\be
	A^i = \theta (x^-) {i \over g} V \nabla^i V^\dagger
\ee
where
\be
	V^\dagger (x_T)= P exp \left( ig \int^\infty_{-\infty} dz^- 
\alpha (z^-,x_T) \right)
\ee

We now have an explicit expression for the gluon field in terms of 
the sources.\cite{k},\cite{mkmw}
For our Gaussian weight function, we can now compute the expectation value
of the gluon fields which gives the gluon distribution function.
The details of such a computation are given in Ref. \cite{mkmw}.
It is a straightforward computation to perform:  One can expand out the 
exponentials and compute term by term in the expansion.  The series 
exponentiates.  One subtlety occurs due to logarithmic infrared infinity 
which is regulated on a scale of order of a Fermi where transverse charge 
correlations go to zero since all hadrons are color singlet.\cite{mah}.
The result is
\be
	< A^i_a (\vec{x},x^+) A^i_a(\vec{0},x^+ ) >
& = & {{N_c^2-1} \over {\pi \alpha_s N_c}} {1 \over x_T^2} \times
\nonumber \\
 & & \left(
1- exp\{ x_T^2 Q_s^2 ln(x_T^2 \Lambda_{QCD}^2)/4\} \right)
\ee
In this equation, the saturation momentum is defined as
\be
	Q_s^2 = 2\pi N_c \alpha_s^2 \int dx^- \mu^2
\ee
and is of the order  $\alpha_s^2$ times the
charge squared per unit area.  

\begin{figure} 
\begin{center} 
\includegraphics[width=0.5\textwidth]
{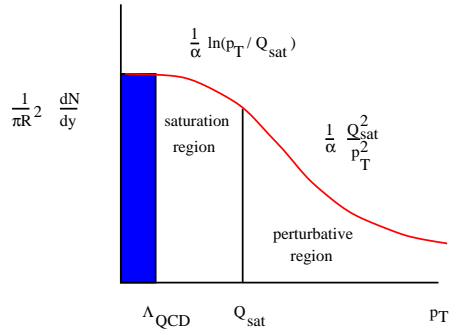} 
\caption{The gluon distribution function. } 
\label{pt} 
\end{center}
\end{figure}

This expression can be Fourier transformed to produce the gluon 
distribution function, with result as shown in Fig. \ref{pt}.
We can understand this plot from the properties of the coordinate space 
distribution function.  We notice that the dominant scale factor in the 
problem is $Q_s$, so to a first approximation everything scales in terms of 
this quantity.  Large $p_T$ corresponds to small $x_T$, and the coordinate 
space distribution behaves as $ln(x_T^2)$ which corresponds to $1/p_T^2$.
This is typical of a bremstrahlung spectrum.  At larger $x_T$, distribution
is of order $1/x_T^2$, which Fourier transforms into $ln(p_T^2)$ at
small $p_T$.  The
softer $x_T$ dependence at large $x_T$ can be traced to a dipole cancelation
of the fields.  The monopole charge field, seen at short distances is 
$ln(x_T^2)$ and the dipole cancelation should set in at large distances when
one cannot resolve individual charges, and reduce this by two powers of $x_T$.
 
The overall scale of the curve is $1/\alpha_s$.  The quantity we are plotting
is in fact the phase space density of gluons.  At small $\alpha_s$, this
density becomes large, and the Color Glass becomes a condensate.  Hence the 
name, Color Glass Condensate.

This form of the gluon distribution function illustrates how the problems
with unitarity can be solved.  Let us assume that the saturation momentum
$Q_s$ is rapidly increasing as $x \rightarrow 0$.  If we start with an $x$ 
so that $Q >> Q_s$, then as $x$ decreases, the number of gluons which
can be seen in scattering rises like $Q_s^2$.  (See Fig. \ref{pt}.) 
Eventually, $Q_s$ becomes 
larger than $Q$, in which case, the number of gluons rise slowly, like 
$ln(Q_s)$.  At this point the cross section saturates since the number of
gluons which can be resolved stops growing, and we are consistent with
unitarity constraints.

The gluon distribution function is defined to be
\be
	xG(x,Q^2) = \int_0^{Q^2} d^2p_T {{dN} \over {d^2p_T dy}}
\ee
This behaves in the saturation region as $\pi R^2 Q^2$, and in the
large $Q$ region as $\pi R^2 Q_s^2$.  We expect that $Q_s^2 \sim charge^2/area$
and due to the random nature of the way charges add, $Q_s^2 \sim R$.  Therefore
in the saturation region, the gluon distribution function is proportional to
the surface area of the hadron, that is the gluons can only be seen which are
on the surface of the hadron.  In the large $Q$ region, one sees gluons 
from the entire hadron, that is, the hadron has become transparent.

\subsection{The Structure of the Gluon Field}

\begin{figure} 
\begin{center} 
\includegraphics[width=0.5\textwidth]
{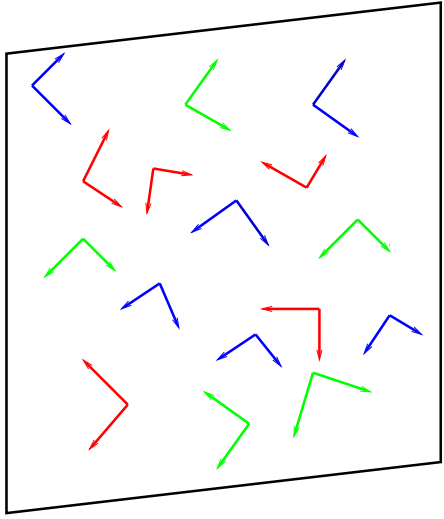} 
\caption{The non-abelian Lienard-Wiechert potentials which form
the Color Glass Condensate. } 
\label{colorglass} 
\end{center}
\end{figure}

The gluon field arises from a charge density which is essentially 
delta function in $x^-$.  In order to solve the equations of motion, the
field must have a discontinuity at $x^- = 0$.  This can be achieved with
a field which is a two dimensional gauge transform of zero field strength
on one side of the sheet and a different gauge transform of zero on the other 
side.  The field strength $F^{\mu \nu}$ is therefore zero if $\mu$ and $\nu$ 
are both in the two dimensional transverse space.  If either index is $-$,
it also vanishes since there is no change in the $x^+$ direction.  The only
non-vanishing component is therefore $F^{i+}$, and this is a delta function
in the $x^-$ direction.  Since $F^{i\pm} = E^i \pm \epsilon^{ij}B^j$,
we see that
\be
	\vec{E} \perp \vec{B} \perp \vec{z}
\ee
The fields are therefore transversely polarized to the direction of motion
and live in the two dimensional sheet where the charges sit.  These
are the non-abelian generalizations of the Lienard-Wiechert potentials of
electrodynamics.
The density
of these fields is of order $1/\alpha_s$.  A picture of the
Color Glass Condensate is shown in Fig. \ref{colorglass}.

\section{Lecture III: Hadron-Hadron Collisions and the Initial Conditions for 
Heavy Ion Collisions}

\subsection{Phenomenology of Mini-Jets}

In the last lecture, we argued that at small x, the typical gluon
constituent of a hadron acquires a transverse momentum of order $Q_s$,
and that this can grow as $x \rightarrow 0$.  This leads us to hope
that in hadron-hadron collisions, this will be the typical momentum scale of 
particle production.  If true, then the processes are weakly coupled and 
computable using weak coupling methods.

This is reminiscent of past attempt to compute particle production by
mini-jets\cite{kaj}-\cite{xnwang}.
On dimensional grounds, the cross section for jet production
$d\sigma /dy d^2p_T \sim \alpha_s^2 /p_T^4$.  If we attempt to compute the
total cross section for jet production,
\be
	{{d\sigma} \over {dy}} \sim \alpha_s^2 \int_{\Lambda^2_{QCD}}
{{d^2p_T} \over p_T^4}
\ee
the result is infrared sensitive, and presumably would be cutoff at 
$\Lambda_{QCD}$.  In early computations, one introduced an ad hoc cutoff
which was fixed, and hopefully large enough so that one could compute
the minijet component.  This of course left unanswered many questions 
about the origin of this cutoff, and the effects of particles produced 
below the cutoff scale.

\begin{figure} 
\begin{center} 
\includegraphics[width=0.5\textwidth]
{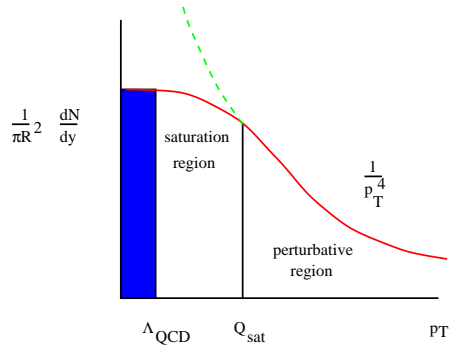} 
\caption{The $p_T$ distribution for mini-jets produced by a Color
Glass Condensate. } 
\label{pt4} 
\end{center}
\end{figure}

In this lecture, we will argue that the Color Glass Condensate cuts off the
integral at a scale of order $Q_s$, the saturation momentum.  At large $p_T$,
dimensional arguments tell us that the density of produced particles has the
form
\be
	{1 \over {\pi R^2}} {{dN} \over {d^2p_Tdy}} = \kappa
{1 \over \alpha_s} {Q_s^4 \over p_T^4}
\ee
The factor of $1/p_T^4$ comes about because the high $p_T$ tail is controlled
by perturbation theory. The $1 /\alpha_s$ arises because of the large density
of gluons in the condensate.
In fact, if we can successfully formulate the
particle production problem classically, we expect that in general
\be
        {1 \over {\pi R^2}} {{dN} \over {d^2p_Tdy}} = {1 \over \alpha_s}
F(Q_s^2/p_T^2)
\label{ptmini}
\ee
At large $p_T >> Q_s$, $F \sim Q_s^4/p_T^4$ and for small $p_T << Q_s$,
$F$ should be slowly varying (logarithmic) or a constant.  A plot is shown
in Fig. \ref{pt4}. 

A word of caution should be injected about the interpretation of mini-jet
production.  Typically it is assumed that there is a simple relationship 
between the multiplicity of produced gluon jets and the multiplicity of pions.
Usually, $N_{pion}$ is taken to be some constant of order one times
$N_{gluon}$.  In our considerations, we can only talk about the gluon 
mini-jet production, and it is beyond the scope of these lectures to relate
this to the final state multiplicity.  Suffice it to say that the situation
is controversial, particularly in heavy ion collisions where there can be
much final state interaction.\cite{son}.

Recall that in heavy ion collisions, we expect that $Q_s^2 \sim A^{1/3}$.
At large $p_T$, Eqn. \ref{ptmini} predicts that
\be
	{{dN} \over {d^2p_Tdy}} \sim \pi R^2 {Q_s^4 \over p_T^4}
\sim {A^{4/3} \over p_T^4}
\ee
This result is consistent with hard incoherent scattering.
At small $p_T$,
\be
	{{dN} \over {d^2p_Tdy}} \sim \pi R^2 \sim A^{2/3}
\ee
which is consistent with much shadowing, and the gluons are produced from
the surface of the nuclei.

The total multiplicity per unit rapidity
\be
	{{dN} \over {dy}} \sim R^2 \int_{Q_s^2} d^2p_T {{Q_s^4} \over p_T^4}
\sim R^2 Q_s^2 \sim A
\ee
is proportional to $A$, just as in color string models!  This is because for 
the Color Glass Condensate, the cutoff in transverse momentum depends on
$A$.  (If one was careful with the factors of $\alpha_s$ in the above 
equation, one would predict mild logarithmic modifications of the linear 
dependence on $A$.) 
In addition to the $A$ dependence, there is also a correlation between
the energy dependence of the gluon distribution function at
saturation  and the multiplicity
of minijets since
\be
	\pi R^2 Q_s^2 =  \int_x^1 dx^\prime G(x^\prime,Q_s)
\ee
a relationship which follows from the last lecture.

We can be a little more careful with the numerical factors which determine the
saturation momentum.\cite{mclgyu}  Using the results of last lecture,
\be
	Q_s^2 = {{2\pi N_c \alpha_s^2} \over {\pi R^2 (N_c^2 -1)}} 
Q_{color}^2
\ee
Here $Q_{color}^2$ is the color charge squared of all quarks and gluons
at larger $x$ values than that of interest.  For a quark,
\be
   Q_{quark}^2 = {1 \over {N_c (N_c^2-1)}} tr ~\tau_a^2 = {1 \over {2N_c}}
\ee
and for a gluon
\be
   Q_{gluon}^2 = {N_c \over {(N_c^2-1)}}   	 
\ee
We find that
\be
	Q_{color}^2 = {N_{quark} \over {2N_c}} +  {{N_c N_{gluon}} \over 
{(N_c^2-1)}}
\ee
If we plug in numbers, at RHIC energies corresponding to 
$x \sim 10^{-2}$, $Q_s \sim 1-2~GeV$

\subsection{Classical Description of Hadron Collisions}

\begin{figure} 
\begin{center} 
\includegraphics[width=0.5\textwidth]
{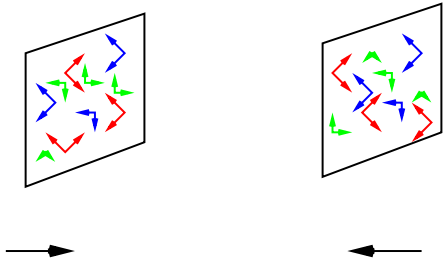} 
\caption{A collision of two ultra-relativistic hadrons. } 
\label{sheetonsheet} 
\end{center}
\end{figure}

We want to describe the collision of two ultra-relativistic hadrons.  
A collision is shown in
Fig. \ref{sheetonsheet}.  The hadrons have been Lorentz contracted to thin 
sheets and a Color Glass Condensate sits in the planes of both sheets.

Before the collision the non-zero fields are for right moving nucleus,
$F^{i+} \sim \delta(x^-) $ and for the left moving nucleus
$F^{i-} \sim \delta(x^+)$.  Before the nuclei pass through one another,
nothing happens and the fields in each sheet are static.  When they 
pass through
one another, the sum of these two fields is not a solution of the equations
of motion, unlike the case in electrodynamics, and this induces a time 
evolution of the fields.\cite{kmw}  

\begin{figure} 
\begin{center} 
\includegraphics[width=0.5\textwidth]
{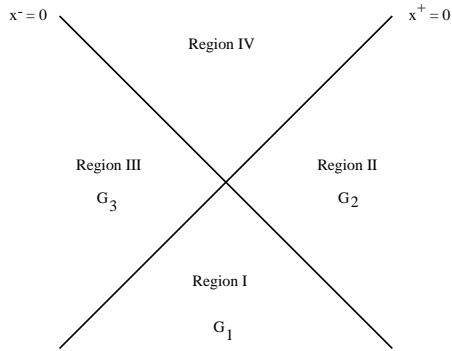} 
\caption{A space-time diagram for the vector potentials in hadron-hadron
scattering. } 
\label{lightcone} 
\end{center}
\end{figure}

One can understand this from the vector potentials.  In Fig. \ref{lightcone}
a space-time diagram is shown for the scattering.
In the backward light cone, Region I, the field is a pure two dimensional
gauge transform of zero field.  In crossing into Regions II and III,
the fields must have a discontinuity to match the charge on the surfaces
of the lightcone.  This requires the vector potential to be different gauge 
transforms of zero field strength, $G_2$ and $G_3$ in these regions.
Now in going to Region IV, one could solve either for the sources on
the left edge of the forward light cone  with a gauge transform of zero or 
the right edge of the
forward light cone with a different gauge transform of zero.
One cannot satisfy the equations of motion for the fields in the presence
of the sources on both edges of the
light cones with the same gauge transform of zero field
strength.  One must produce a field in the forward light cone which is not
a gauge, and therefore matter is produced.

The situation in QCD is completely different than that in electrodynamics.
In electrodynamics, one must produce pairs of charged particles to make matter
in the forward light cone.  This  arises from a quantum correction
to the equations of motion.  In QCD, matter is produced classically.

The procedure for solving this problem is now straightforward, in principle.
One solves the classical equations of motion in the forward light cone with 
boundary conditions at the edges of the forward light cone.  We will shortly
determine these boundary conditions and the form of the solution in the forward
light cone.  Then one evolves the equations of motion into the far future.
At some time, the energy density becomes dilute, and the field equations
should linearize in some gauge.  One can then identify the quanta of the
linearized fields in the standard way that one does classical radiation theory
in electrodynamics.

\subsection{The Form of the Classical Field}

Before the collision, the form of the classical field can be taken as
\be
	A^+ & = & A^- = 0 \nonumber \\
        A^i & = & \theta (x^-) \theta (-x^+) \alpha^i_1 (x_T)
             +\theta (-x^-) \theta (x^+) \alpha^i_2 (x_T)
\ee
where the $\alpha^i$ are two dimensional gauge transforms of zero
field.  We will consider the collision of identical hadrons.  The solution
in the forward light cone is therefore expected to be boost invariant.
After the collision, a boost invariant solution is
\be
	A^+ & = & x^+ \alpha(\tau, x_T) \nonumber \\
        A^- & = & x^- \beta(\tau, x_T) \nonumber \\
        A^i & = & \alpha^i_3(\tau, x_T)
\ee
We can choose the gauge
\be
	x^+ A^- + x^- A^+ = 0
\ee
so that 
\be
	\alpha(\tau, x_T) = - \beta (\tau, x_T)
\ee

In the forward light cone, the equations of motion are
\be
	{ 1 \over \tau^3} \partial_\tau \tau^3 \partial_\tau \alpha
- [D^i, [D^i, \alpha]] = 0
\ee
and
\be
	{1 \over \tau} \partial_\tau \tau \partial_\tau \alpha^i_3
- ig \tau^2 [\alpha,[D^i, \alpha]] - [D^j, F^{ji}] = 0
\ee

The boundary conditions are determined by matching the solution in Regions
II and III to that in the forward light cone.  The result is
that $\alpha$ and $\alpha^i_3$ must both be regular as $\tau \rightarrow 0$
and
\be
	\alpha_3^i (0,x_T) & = & \alpha_1^i(x_T) + \alpha_2^i (x_T)
\nonumber \\
        \alpha(0,x_T) & = & 
{{-ig} \over 2} [\alpha^i_1 (x_T), \alpha_2^i (x_T)]
\ee
The problem is now well defined, and these equations may be numerically solved.

The behaviour of these solution at large $\tau$ can be extracted.  With
$V(x_T)$ an element of the group, the solution is a small fluctuation field
up to a possible large gauge transformation
\be
	\alpha (\tau, x_T) & = & V \epsilon(\tau , x_T ) V^\dagger \nonumber \\
        \alpha^i_3 (\tau, x_T) & = & V (\epsilon^i_3(\tau,x_T) + {i \over g}
\partial^i )V^\dagger
\ee
The small fluctuation fields $\epsilon$ and $\epsilon^i$ solve the equations
\be
	{1 \over \tau^3} \partial_\tau \tau^3 \partial_\tau \epsilon -
\nabla^2_T \epsilon = 0
\ee
and
\be
         {1 \over \tau} \partial_\tau \tau \partial_\tau 
\epsilon^i - (\nabla^2_T \delta^{ij}
- \nabla^i \nabla^j) \epsilon^j = 0
\ee
At large $\tau$, these linear equations can be Fourier analyzed with the result
\be
	\epsilon^a (\tau, x_T ) = \int 
{{d^2k_T} \over {(2\pi )^2 \sqrt{2 \omega}}} 
{1 \over \tau^{3/2}} \left( a_1^a(k_T) e^{-ik \cdot x} + c. c. \right)
\ee
and
\be
        	\epsilon^{ai} (\tau, x_T ) = \int 
{{d^2k_T} \over {(2\pi )^2 \sqrt{2 \omega}}} \epsilon^{ij} {k^j \over \omega}
{1 \over \tau^{1/2}} \left( a_2^a(k_T) e^{-ik \cdot x} + c. c. \right)
\ee
In these equations $\omega = \mid k_T \mid $.

One can compute the energy distribution associated with these fields as
\be
	{{dE} \over {dy d^2k_T}} = {\omega \over {(2\pi)^2}} \sum_{ia} 
\mid a_i^a (k_T ) \mid^2
\ee
and the multiplicity distribution is given by dividing this by $\omega$, that 
is
\be
{{dN} \over {dy d^2k_T}} = {1 \over {(2\pi)^2}} \sum_{ia} 
\mid a_i^a (k_T ) \mid^2
\ee	 
These last two formulae correspond to those of free quantum
filed theory when we replace $a(p), a^*(p)$ by the creation
and annihilation operators $a(p), a^\dagger (p)$. 
The $a(p)$ are the classical quantities which correspond to
the quantum creation and annihilation operators.  These formulae show
how to use the classical solutions to compute distributions of produced 
minijets.
 
\subsection{Numerical Results for Mini-Jet Production}

Krasnitz and Venugopalan have numerically solved the classical equations
for mini-jet production.\cite{kv}  
This involves finding a gauge invariant
discretization of the classical equations of motion.  One then
solves the classical equations for a fixed $\rho_1$ and $\rho_2$, and 
extracts the produced radiation.  An ensemble of sources are produced with
the Gaussian weight of the Color Glass, which then produces an ensemble
of radiation fields.  These fields are then averaged to generate the mini-jet
distributions.

\begin{figure} 
\begin{center} 
\includegraphics[width=0.5\textwidth]
{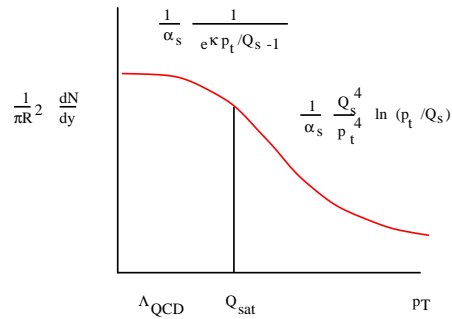} 
\caption{An illustration of the results generated by numerical simulation
of the classical equations for mini-jet production. } 
\label{ptraju} 
\end{center}
\end{figure}

In Fig. \ref{ptraju}, the form of the numerical results for mini-jet production
is illustrated.  At large $p_T$, the results of analytic studies are 
reproduced which up to logarithms is $\sim 1/p^4_T$.  At $p_T \le Q_s$,
the distribution flattens out.\cite{kmw}-\cite{mclgyu}   
To good numerical accuracy, the result in this region can be fit
to a two dimensional Bose Einstein distribution,
\be
	{1 \over {\pi R^2}} {{dN} \over {d^2p_Tdy}} \sim {1 \over \alpha_s}
{1 \over {e^{\kappa p_T/Q_s} -1}}
\ee
where $\kappa$ is a constant of order 1.  

The result at large $p_T$ can be computed analytically by expanding the
equations in powers of the gluon field.  At high $p_T$, the phase space is 
not so heavily occupied, so a field strength expansion makes sense.
At small $p_T$, it is not at all certain that the result is in fact
an exact two dimensional Bose-Einstein 
distribution.\cite{kv}-\cite{ksea}  In any case, 
the origins of this distribution have nothing at all to do with thermodynamics,
and it is a useful example of the traps one can fall into if one
assumes that exponential distribution corresponds to a temperature and 
thermalization.

\subsection{pA Scattering}

\begin{figure} 
\begin{center} 
\includegraphics[width=0.5\textwidth]
{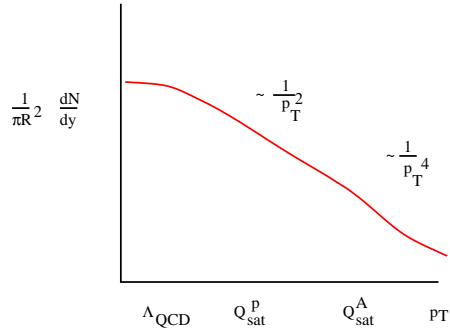} 
\caption{The $p_T$ distribution for particles produced in a $pA$ collision. } 
\label{pa} 
\end{center}
\end{figure}

An interesting example of minijet production is given by the collision
of two hadrons of different size.\cite{kovmu}-\cite{dumcl}  
We will generically refer to this as $pA$
scattering, although most of our considerations could be generalized to 
$A^\prime A$ nuclear collisions.   In Fig. \ref{pa},
the transverse momentum distribution for $pA$ scattering is shown.

There are three distinct regions, which follow
from the fact that there are two saturation scales, $Q_s^A$ and $Q_s^p$,
and $Q_s^p << Q_s^A$ since $(Q_s^A)^2 \sim A^{1/3}$.
At very large $p_T$ where the fields from
both nuclei are small, the distribution can be computed from perturbation 
theory, and the distribution falls as $1/p_T^4$ and is proportional to
$(Q^A_s Q^p_s)^2$.  This first region is for $p_T >> Q_s^A$.
An intermediate region where the field from the nucleus
is strong but the field from the proton is weak and can be treated 
perturbatively.  This intermediate region is for $Q_s^p << p_T << Q_s^A$.
There is finally the region where $p_T << Q_s^p$ where both fields are strong.

We expect that in the intermediate region, the transverse momentum 
dependence will be in between the flat behaviour at small $p_T$ and the 
$1/p_T^4$ behaviour characteristic of large $p_T$,  The naive expectation is 
$1/p_T^2$ in the intermediate region.  The total multiplicity can be
computed if one understands this intermediate region since the dominant
contribution arises here.  The strength in this intermediate region should
involve the total charge squared from the proton, but that from the nucleus
should go like $p_T^2$ so that when combined with the $1/p_T^4$, one gets
a distribution proportional to $1/p_T^2$.   This softer behaviour of the 
distribution function follows since we are inside the region where we 
expect coherence from the field of the nucleus, and since the distribution
should extrapolate between $1/p_T^4$ at very large $p_T$ and a constant,
up to logarithms, at very small $p_T$.

In fact, it is possible to compute the behaviour in this intermediate region.
The equations for the classical production can be analytically solved for any
$p_T >> Q_s^A$.  The solution in the forward light cone are plane waves
which are gauge 
transformed by the field of the large nucleus.  
The boundary conditions determine the 
strength of these waves.

\begin{figure} 
\begin{center} 
\includegraphics[width=0.5\textwidth]
{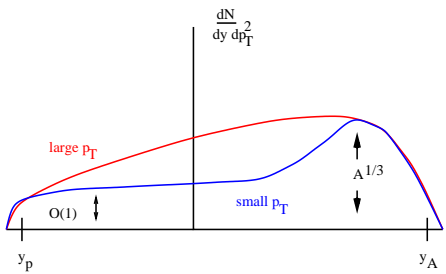} 
\caption{The distribution in rapidity for $dN/dyd^2p_T$ in a $pA$ collision. } 
\label{ya} 
\end{center}
\end{figure}

For the total multiplicity, in the large $p_T$ region $dN/dyd^2p_T
\sim A^{1/3}$.  We expect that as we interpolate between the proton 
fragmentation region and that of the nucleus, we go between $O(1)$ and
$O(A^{1/3})$ as shown in Fig. \ref{ya}.
For $p_T$ in the intermediate region, we expect that $dN/dyd^2p_T$ is of order
1 except for a small region of rapidity around the fragmentation region of the 
nucleus.  The total integrated multiplicity arises from this latter region
so we expect that $dN/dy \sim O(1)$.

\subsection{Thermalization}

After the gluons are produced in hadron-hadron collisions, 
they may rescatter from one another.\cite{son}  If one goes
to very small x so that the density of gluons becomes very large, one expects 
that the gluons will eventually thermalize.  Due to the very large typical
$p_T$, $\alpha_s << 1$, and this takes a time $\tau \sim 1/(\alpha_s^2 Q_s)$ 
which is longer by a factor of $1/\alpha_s^2$ than the natural time scale.
The system therefore becomes  dilute relative to its natural scale.

\begin{figure} 
\begin{center} 
\includegraphics[width=0.5\textwidth]
{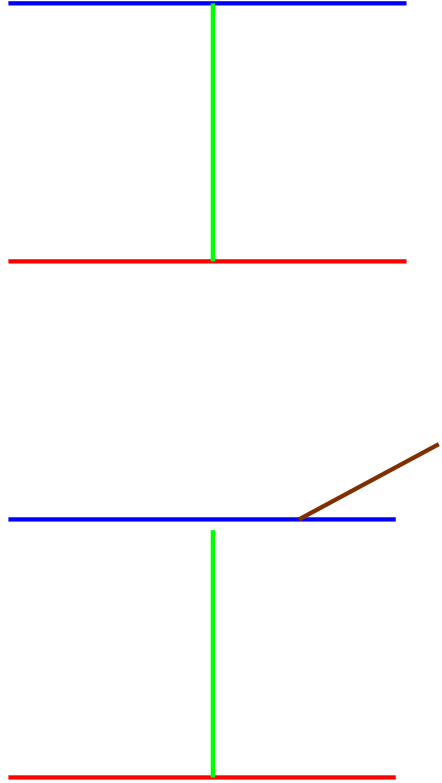} 
\caption{The diagrams for gluon scattering which lead to thermalization. } 
\label{gluescat} 
\end{center}
\end{figure}

In the first diagram of Fig. \ref{gluescat}, there is ordinary Coulomb 
scattering.  When all processes which populate and
depopulate phase space are summed, this diagram is only naively logarithmically
divergent, and is cutoff by the density dependent Debye mass.
$\rho_{gluons} << p_T^3$.
In the second diagram, there is no such cancellation, and the diagram is of 
order $1/(\alpha_s \sqrt{\rho_{gluons}})$.  At a time of order
$1/(\alpha_s^2 Q_s)$ for a density decreasing like $1/\tau$ as we expect for
ultrarelativistic nuclear collisions, the diagram is enhanced by a factor
of $1/\alpha_s$.  This cancels the extra factor of $\alpha_s$ coming from 
the diagram being higher order in perturbation theory.\cite{son}.

What appears to happen is that as the system gets more dilute, it thermalizes 
due to multigluon production.  This will modify the relationship between the
number of gluons produced as mini-jets and the pion multiplicity.  
How this actually works is not yet fully understood.

\section{Lecture 4:  The Renormalization Group}

The effective action for the theory we have described must be gauge
invariant and properly describe the dynamics in the presence of external
sources.  For the theory which we have written down in past lectures
with the $J \cdot A$ coupling of source to field, gauge invariance is only
retained if we impose 
\be
        D_\mu J^\mu = 0
\ee
This equation presents a problem in our formulation since it implies that the
source cannot be independently specified from the field.  This did not 
present a problem for the classical theory since one could
find a solution which solved the constraint.  When we compute quantum 
corrections and proceed to a renormalization group treatment, we must be more
careful.

In a clever series
of papers,\cite{mklw} it was shown that one can generalize the $J \cdot A$
coupling.  This lead Jalilian-Marian, Kovner, Leonidov and Weigert
to propose the action
\be
        S & = & - {1 \over 4} \int d^4x F_{\mu \nu}^aF_a^{\mu \nu} + 
 {i \over N_c} \int d^2x_t dx^- \delta (x^-) \nonumber \\ 
& & \times \rho^a(x_t) tr T^a
exp\left\{ i \int _{-\infty}^\infty dx^+ T \cdot A^-(x) \right\}
\ee
In this equation, the matrix $T$ is in the adjoint representation of the
gauge group.  This is required for reality of the action.  When this 
action  is extremized to get the Yang-Mills equations, one can identify
the current and show that the current is covariantly conserved. This action
is invariant under gauge transformations which are identified at $x^+
 = \pm \infty$.  (Even this can be corrected to get a fully invariant theory
if one generalizes even further to complex time Keldysh contours.  As shown
in Ref. \cite{ilm}, this further generalization does not affect the
renormalization group in lowest non-trivial order.) 
This is a consequence of the
gauge invariance of the measure of integration over the sources $\rho$. 
This will be taken as a boundary condition on the theory.   In general
if we had not integrated over sources, one could not define a gauge
invariant theory with a  source, 
as gauge rotations would change the definition of the
source.  Here because the source is integrated over in a gauge invariant
way, the problem does not arise.

In the most general gauge invariant theory which we can write down
is generated from
\be
        Z = \int [d \rho ] e^{-F[\rho ]} \int [dA] e^{iS[A,\rho ]}
\ee
This is a generalization of the Gaussian ansatz described in the
previous lecture.  It allows for a slightly more complicated structure
of stochastic variation of the sources.  The Gaussian ansatz can be
shown to be valid when evaluating structure functions at large
transverse momenta.
\be
        F_{Gaussian} [\rho ] = { 1 \over 2 } \int dx^- d^2 x_t~ 
{{\rho^2 (x_t)}
\over {\mu^2(x^-)}}
\ee

This theory is an effective theory
valid only in a limited range of rapidity much less than the rapidity 
of the source.  The sources for this theory sit at higher rapidity.
This happens because as we go to lower values of rapidity, the fluctuations
in the field are integrated out and are replaced by sources and an integral
over fluctuations in the source.  The renormalization group equations
which we will describe are what make the theory independent of this
cutoff.
To fully determine $F$ in the above equation demands a full solution of
these renormalization group equations.  This has yet to be
done, although there are now approximate solutions for small
and large transverse momentum of the fields.\cite{ilm}-\cite{im}  

We can understand this a little better by imagining what happens when we
compute a quantum correction to the classical theory.  This quantum correction
will generate terms proportional to $\alpha_s ln(\Lambda^+/p^+)$ where
$\Lambda^+$ is the $p^+$ cutoff for our effective theory.  Clearly these
corrections are small and sensible only if 
$e^{-1/\alpha_s} \Lambda^+ << p^+ << \Lambda^+$.  
If we want to generate a good 
effective theory at smaller values of $p^+$, we need to break the theory into
intervals of $p^+$ with each interval sufficiently small so that the quantum 
fluctuations are small and computable.  The relation between one interval
and the next is the renormalization group.

The remarkable thing that happens when one integrates out the fluctuations 
interval is that only the function $F$ which controls the source strength is 
modified!   The functional form of $F$ is modified so that this equation is of
the form
\be
	{d \over {dy}} Z = -H(\rho, \delta/\delta\rho) Z
\label{rgh}
\ee
where
\be
	y = ln(\Lambda^+_i/\Lambda^+_f)
\ee
and $\Lambda^+_{i,f}$ are the cutoffs at  the initial and final values,
and
\be
	Z = e^{-F}
\ee
This equation is of the form of the time evolution for a two spatial dimension
quantum field theory where the coordinates are $\rho $ and the momenta
are $d/d\rho$.

\subsection{How to Compute the RG Effective Hamiltonian}

In the Eqn. \ref{rgh},  the renormalization group Hamiltonian H was introduced.
I will here outline how it is computed.  We first take the theory
defined for $p^+ < \Lambda^+_i$.  We integrate out the quantum fluctuations.
In particular, the two point function is
\be
	{\cal G}^{ij} (x,y) = < (A^i(x) + \delta A^i(x))
(A^i(y) + \delta A^i(y))>
\ee
In this equation, $A$ is the classical background field and
$\delta A$ is the small fluctuation.
At the momentum scales which will be of interest for $p^+ < \Lambda^+_f$, it 
is sufficient to consider the equal time limit of this correlation function.
We now identify
\be
 < \delta A \delta A > & = & G <\delta \rho \delta \rho > G \nonumber \\
 & = & G \chi G
\ee
where $G$ is the Greens function in the classical background field $A$.
We also identify
\be
	<\delta A > & = & G < \delta \rho > \nonumber \\
 & = & \sigma
\ee
We can get exactly the same result by modifying the weight
function so that we reproduce $\chi $ and $\sigma$ and move the cutoff to
$\Lambda^+_f$ so that there are no longer quantum fluctuations to integrate 
out.  This is the origin of the form of Eqn. \ref{rgh}

Some technical comments about the computations are required.  One must
be extremely careful of gauge.  The gauge prescriptions of retarded or 
advanced for $1/k^+$ singularities are used.  We were not able to effectively
use either Leibbrandt-Mandelstam or principle value prescriptions although
this may be possible in principle.  When one computes propagators in
background fields, one gets analytic expressions in terms of line ordered
phases of the source $\rho$.  It is most convenient to compute these in
$\delta A^- = 0 $ gauge and express things in terms of the source in 
$A^- = 0$ gauge, and then rotate results back to lightcone gauge. This can be 
carefully done only when the $1/k^+$ singularity is properly regularized.

If we change variable to space-time rapidity, we can define
\be
	\alpha(y, x_T) = {1 \over {-\nabla_T^2}}\rho (y,x_T)
\ee
and
\be
	V^\dagger (y,x_T) = P exp\left( ig \int_{-\infty}^y dy^\prime
\alpha(y^\prime,x_T) \right)
\ee
After much work, one finds
\be
	H = {\alpha_s \over 2} \int d^2x_T J^{ia}(x_T) J^{ia}(x_T)
\ee
where
\be
	J^{ia} (x_T) = \int {{d^2z_T} \over \pi } {{(x-z)^i} \over
{(x-z)^2}} (1 -V^\dagger (y,x_T)V(y,z_T))^{ab} {1 \over i} {\delta \over
{\delta \alpha^b(y,z_T)}}
\ee
The Hamiltonian is positive definite and looks like a pure kinetic energy term
(up to the multiplicative non-linearities) with no potential.\cite{ilm}

The renormalization group above can also be seen to be a consequence
of equations for correlation functions of $V(y,x_T)$.\cite{bal}
In Ref. \cite{kmiw} it was shown that these equations for correlation functions
were almost the same as those of the renormalization group.  There was an error
in this analysis associated with the subtleties of gauge fixing, and when
repaired gives that these equations are precisely equivalent.\cite{ilm}.
Meanwhile,  Weigert  showed that the equations for the correlation functions
could be summarized as a Hamiltonian equation of the form above.\cite{weig}
which was also shown to be precisely the equations for the renormalization 
group Hamiltonian.\cite{ilm}

\subsection{Quantum Diffusion}

The Hamiltonian presented in the previous section is analogous to that
without a potential.  If we were to ignore the
non-linearities associated with the matrices $V$, this would
be the Hamiltonian for a free theory with only momenta and no potential.

If there was a potential in the Hamiltonian, then at large times,
the solution of the above renormalization group equations would be trivial,
\be
	Z \sim exp(-y E_o )
\ee 
where $E_o$ is the ground state energy.  All expectation values would become
rapidity independent and the solution to the small $x$ problem would be
trivial:  $x$ independence.

The solution to the above equation is more complicated.  One can see this by
studying a one dimensional quantum mechanics problem:
\be
	{d \over {dy}} Z = - {p^2 \over 2} Z
\ee
with solution
\be
	Z = {1 \over \sqrt{2\pi y}} exp(-x^2/2y)
\ee 
This equation describes diffusion.  The width of the Gaussian in $x$
grows with time.  This is unlike solving
\be
	{d \over {dy}} Z = -\left( {p^2 \over 2} + V(x) \right) Z
\ee
In this latter case, the coordinate $x$ evolves towards the minimum of
$V$, and then does undergo small fluctuations around this minimum.

We see therefore that the non-triviality of the small x problem in QCD 
arises because of the quantum diffusive nature of the renormalization
group equations.

\subsection{Some Generic Features of the Renormalization Group Equation}

If we compute the correlation function of two sources
using Eqn. \ref{rgh}, we find that
\be
	{d \over {dy}} <\rho(x) \rho (y) > = -< \rho(x) \rho(y) H >
\ee
At large $k_T$ when the fields are linear,
the
gluon structure function is the same as the source-source correlation function
up to a trivial factor of $1/k_T^2$.  (The momenta $k_T$ is conjugate to
the coordinate $x_T-y_T$.)  If we ignore the non-linearities in $H$, keeping
the lowest order non-vanishing terms, and if we integrate by parts the
factors of $\delta /\delta \alpha(y,x_T)$, we get a closed linear 
equation for the correlation function.  This is precisely
the BFKL equation.

In fact, in the region where the equations are linear, 
one is in the high $k_T$ limit, and this also reduces to the 
DGLAP and BFKL equations,
which are known to be equivalent if one computes distribution functions to 
leading order in $\alpha_s ln(1/x)ln(Q^2)$, where $Q$ is some typical momentum
for the correlation function, $Q \sim k_T$.  
When the non-linearities are important, the
non-linearities of this equation cannot be ignored.

\begin{figure} 
\begin{center} 
\includegraphics[width=0.5\textwidth]
{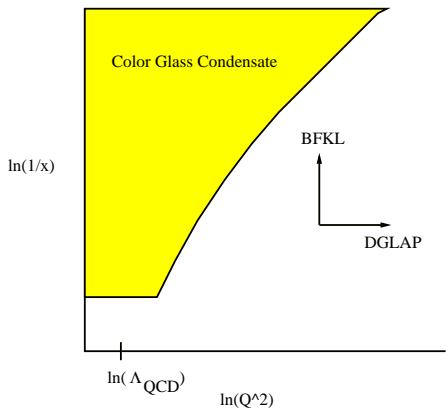} 
\caption{The various regions of evolution for structure functions in the
$ln(1/x)$-$ln(Q^2)$ plane. } 
\label{xq} 
\end{center}
\end{figure}

The situation is as shown in Fig. \ref{xq}.
In the linear region, one can choose to evolve using linear equations.
In the $ln(Q^2)$ direction, the equation is the DGLAP equation and in the
$ln(1/x)$ direction, it is the BFKL equation.  There is a boundary
region in the $ln(1/x)$-$ln(Q^2)$ plane.  Within this boundary region,
there is a high density of glue and the evolution becomes non-linear.
One always collides with this region if one decreases $x$ and holds $Q^2$ fixed
or decreases $Q^2$ holding $x$ fixed.

\subsection{Some Limiting Solutions of the Renormalization Group Equations}

In the small $k_T$ region, we expect that correlation functions such 
as \linebreak
$<V(x) V^\dagger (y)>$ are very small, since we are probing the theory
at distance scales long compared to natural correlation lengths.  
In this limit, one might be able to ignore the non-linearities in the 
renormalization group equations.  Using that $x^i/x^2 = \nabla^i /\nabla^2$,
we have then
\be
	\int d^2x_T~J^2(x_T) \sim \int d^2z d^2z^\prime <z \mid
{1 \over {-\nabla_T^2}} \mid z^\prime > {\delta \over {\delta \alpha (y,z)}}
{\delta \over {\delta \alpha (y,z^\prime)}}
\ee
The solution for Eqn. \ref{rgh} is
\be
	F = {\kappa \over {2\alpha_s}} \int dyd^2x_T \nabla_T^i \alpha (y,x_T) 
\nabla_T^i \alpha (y,x_T)
\ee
The small $k_T$ functional $F$ is a pure scale-invariant Gaussian.  
It is universal and independent of initial conditions.

In the large $p_T$ region, we perform a mean field analysis.  The result
is that discussed in Lecture II.  

For details of the analysis leading to the results of this section, the 
interested reader is referred to \cite{im}.

\subsection{Some Speculative Remarks} 

The form of the renormalization group equation appears to be simple.  
It looks like it might even be possible to find exact solutions.  In  
remarkable works,\cite{bal},\cite{k2} Balitsky and Kovchegov
have shown that the equation for the two line correlation function
$W(x) W^\dagger (y)$ where $W$ is in the fundamental representation
becomes a closed non-linear equation in the large $N_{color}$ limit.
This means that at large $N_{color}$ one can compute this correlation function 
at arbitrarily small x including all the non-linearities associated
with small x.

Although Kovchegov's original derivation was for large nuclei, the result
can be shown to follow directly from the renormalization group Eqn. \ref{rgh}.
This is done by taking the expectation value of $<W(x) W^\dagger (y)>$,
using the form of the Hamiltonian and a factorization property of expectation 
values true in large $N_{color}$.  A derivation is presented in Ref. 
\cite{im}
for the interested reader.

This result is interesting in itself since it means that all of
the saturation effects for $F_2(x,Q^2)$ may be computed at small $x$.  The
Balitsky-Kovchegov equation has been solved numerically.\cite{braun}

More important, it suggests that perhaps, at least in large $N_{color}$,
the full renormalization group equations may be solved for $F$.

\section{Concluding Comments}

In these lectures, constraints of space and time have forced me to
not mention many of the exciting areas that are currently under study.
One of these areas, is diffraction.\cite{buchmuller}-\cite{kovmcl}.  
One can show that the same formalism which gives deep inelastic scattering
also gives diffraction and that there is a simple relation between 
diffractive structure functions and deep inelastic scattering.
I have also not developed a formal treatment of 
deep inelastic scattering within the Color Glass Condensate 
picture.\cite{mueller}-\cite{venmcl}.

The last lecture is very sketchy, and should provide an introduction to the 
literature on this problem.  The derivation of the results discussed in that
lecture are onerous, and all the details have been omitted in these lectures.
In some sense this is good, since the most interesting part of this
problem is to understand and solve the renormalization group equations,
and at least this problem is clearly stated, and free from the technical
details from which it arises..

An area which should be better understood from the perspective presented 
above is the nature of shadowing for nuclei at small x.  This relates
deep inelastic scattering and diffraction in a non-trivial way, and 
the Color Glass Condensate is one of the few theories available which 
pretends to treat both consistently.

The other area where  there is much potential is the production of quarks
in hadron-hadron collisions.  In particular, the charm quark may provide us
a real clue about non trivial dynamics since its mass is very close to the
scale of the Color Glass Condensate for large nuclei at accessible energies.

\section{Acknowledgments}

These lectures were delivered at the
40'th Schladming Winter School: DENSE MATTER, 
March 3-10, 2001.  I thank the organizers for making this a very informative
meeting, with an excellent atmosphere for interaction. 

I thank my colleagues Alejandro Ayala-Mercado, Adrian Dumitru, Elena Ferreiro,
Miklos Gyulassy, Edmond Iancu
Yuri Kovchegov, Alex Kovner, Jamal Jalilian-Marian, Andrei Leonidov,
Raju Venugopalan and Heribert Weigert with whom the ideas presented
in this talk were developed.  I particularly thank Kazunori Itakura for
a careful and critical reading of the manuscript.

This manuscript has been authorized under Contract No. DE-AC02-98H10886 with
the U. S. Department of Energy.

\end{document}